\documentclass[letter,traditabstract]{aa}
\usepackage{amssymb}
\usepackage{graphics}
\usepackage{tabularx}
\usepackage{color}
\usepackage{natbib}
\usepackage{rotating}

\def\css{C/2013\,A1}

\begin{document}


\title{Comet C/2013\,A1 (Siding Spring) as seen with the \\
Herschel Space Observatory\thanks{Herschel is an ESA space 
observatory with science instruments provided by European-lead Principal
Investigator consortia and with important participation from NASA.}}

\author{%
Cs. Kiss\inst{1} \and
T.G. M\"uller\inst{2} \and
M. Kidger\inst{3} \and
P. Mattisson\inst{4} \and
G. Marton\inst{1}}
\institute{%
Konkoly Observatory, MTA Research Centre for Astronomy and Earth Sciences,
	Konkoly Thege 15-17,
	1121 Budapest, Hungary; e-mail: \texttt{kiss.csaba@csfk.mta.hu} 
\and 
Max-Planck-Institut f\"ur extraterrestrische Physik, 
	Postfach 1312, Giessenbachstr., 85741 Garching, Germany 
\and
Herschel Science Centre, ESAC, European Space Agency, 
28691 Villanueva de la Ca\~nada, Madrid, Spain
\and
STAR, Stockholm Amateur Astronomers, Drottninggatan 120, 11360 Stockholm, Sweden
}


\date{Received \dots; accepted \dots}

\abstract{The thermal emission of comet C/2013\,A1 (Siding Spring) was observed 
on March 31, 2013, at a heliocentric distance of 6.48\,au
using the PACS photometer camera of the Herschel Space Observatory. 
The comet was clearly active, showing a coma that could be traced to a distance 
of $\sim$10\arcsec{}, i.e. $\sim$50000\,km. Analysis of the
radial intensity profiles of the coma provided dust mass and dust production
rate; the derived grain size distribution characteristics indicate an
overabundance of large grains in the thermal emission. We estimate that 
activity started about 6 months before these observations, 
at a heliocentric distance of $\sim$8\,au. }


\maketitle


\section{Introduction}

Comet C/2013\,A1 (Siding Spring) will approach Mars to 140\,000\,km on October 19, 2014
and may cause a notable fluence of large, high velocity particles, posing a threat on 
instruments working either on the planet's surface or in Martian orbits. 
Several authors performed hazard analysis based 
on various datasets \citep{Farnocchia,Kelley,Tricarico,Ye} and in most cases 
obtained a low fluence of dust particles at the critical places. 
In this letter we report on the Herschel/PACS observations of the comet, 
analyse the coma structure and construct simple models to estimate the 
dust production rate, grain size distribution and onset time of the comet's activity. 
These far-infrared observations are particulary well suited
to complement previous observations used for dust particle impact calculations, 
as they were obtained 1.5 years before the
encounter. Particles visible at the time of these infrared observations have
the chance to reach the surface or close orbits around Mars \citep{Kelley}, while
large dust grains ejected later will likely miss the surface of the red planet. 

\section{Observations and data reduction}

Thermal emission of \css{} was observed
with the PACS photometer camera 
\citep{poglitsch2010} of the Herschel Space Observatory \citep{pilbratt2010}
using the time awarded in a DDT proposal exclusively for \css{} 
(proposal ID: DDT\_pmattiss\_1, P.I.: P.~Mattisson). The observations were performed in
mini-scanmap mode and covered all three bands using the configurations detailed 
in Table~\ref{table:herschelobs}. 
At the time of the observations
the target was at a heliocentric distance of r\,=\,6.479\,au, 
at a distance of $\Delta$\,=\,6.871\,au from Herschel, and at a 
phase angle of $\alpha$\,=\,7\fdg98. 

The data reduction is based on the pipeline
developed for the "TNOs are Cool!" Herschel Open Time Key Program
\citep{mueller2009,Kiss2014}. 
The reduction of raw data was performed using an optimized version of the PACS 
bright point source pipeline script with the application of 
proper motion correction, i.e. the maps have been produced in the co-moving
frame of the comet.
While the movement of the target was significant, it did not move enough 
between two OBSIDs that the maps could be used as mutual 
backgrounds. Our images may therefore be affected by 
background features.

Maps were created with both applying  
high-pass filtering (HPF) in combination with the \emph{photProject()} task 
in HIPE, as well as using the standard JScanam pipeline. 
The HPF+photProject maps are better suited for point- and compact sources,
since due to the high-pass filtering the large scale structure (extended
emission) is not preserved. In contrast, JScanam maps keep the larger
scale structures in the maps.
The typical spatial scale on which extended emission 
is suppressed in HPF maps is $>$30\arcsec with our settings. 
The comparison of the HPF and JScanam maps show that they are practically
identical in the central $\sim$30\arcsec area, and 
no additional extended emission could be identified at these  
distances on the JScanam maps. In addition, the HPF maps
provided a significantly grater S/N ratio in the red band than 
the JScanam maps, therefore we used these HPF images for further analysis
(Fig.~\ref{fig:maps}). 

\begin{table}
\caption{Summary of Herschel observations.
The columns are: 
i) observation identifier (OBSID); 
ii) date and start time;
iii) duration; 
v) filters configuration;
vi) scan angle direction with respect to the detector array.}
\begin{center}\begin{tabular}{cccccr}
\hline
 OBSID & Date    & t$_{obs}$ & Filters          & p$_{scan}$ \\
       & \& time & (s)      & ($\mu$m/$\mu$m)   & (deg) \\            
\hline
            &   2013-03-31 &     &          &       \\
 1342268974 & 	18:10:51 &  1132 &	70/160	&	70	\\
 1342268975 & 	18:30:46 &  1132 &  70/160	&	110	\\
 1342268976 & 	18:50:41 &  1132 & 	100/160	&	70	\\
 1342268977 & 	19:10:36 &  1132 &	100/160	&	110	\\
\hline

\hline
\end{tabular}\end{center}\vspace*{-3mm}
\label{table:herschelobs}
\end{table}

\begin{figure*}
\hbox{\includegraphics[width=6cm]{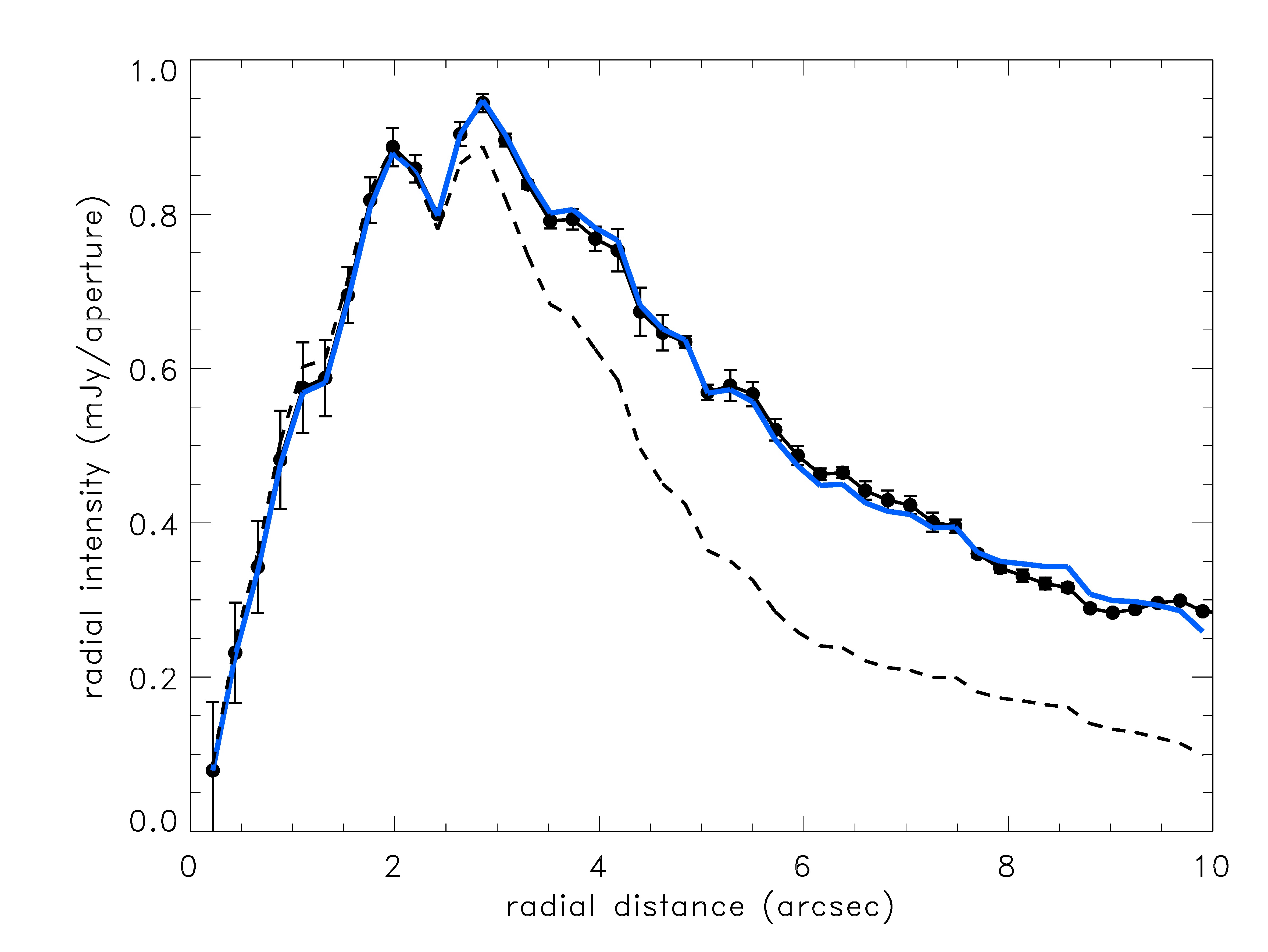}
\includegraphics[width=6cm]{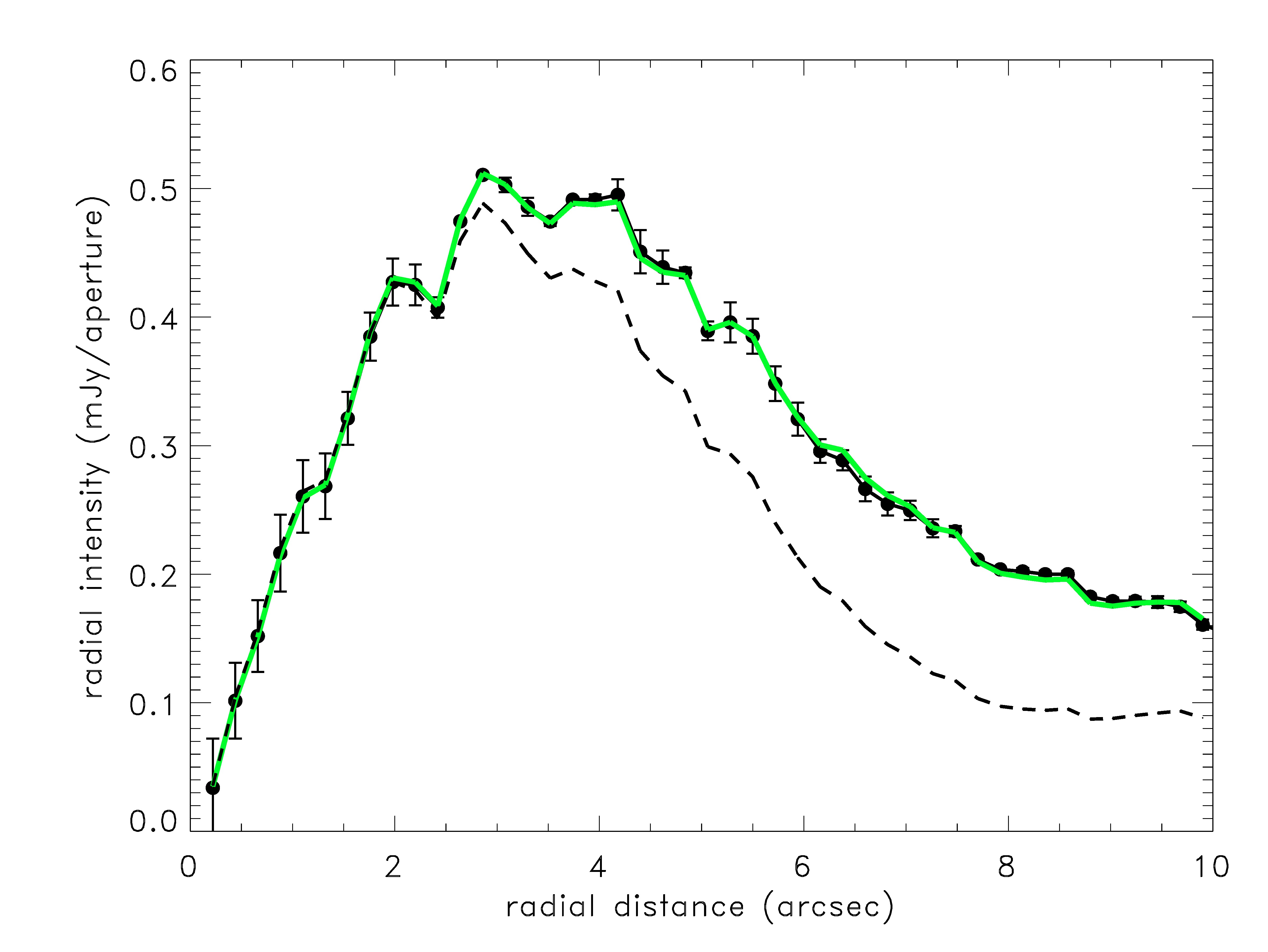}
\includegraphics[width=6cm]{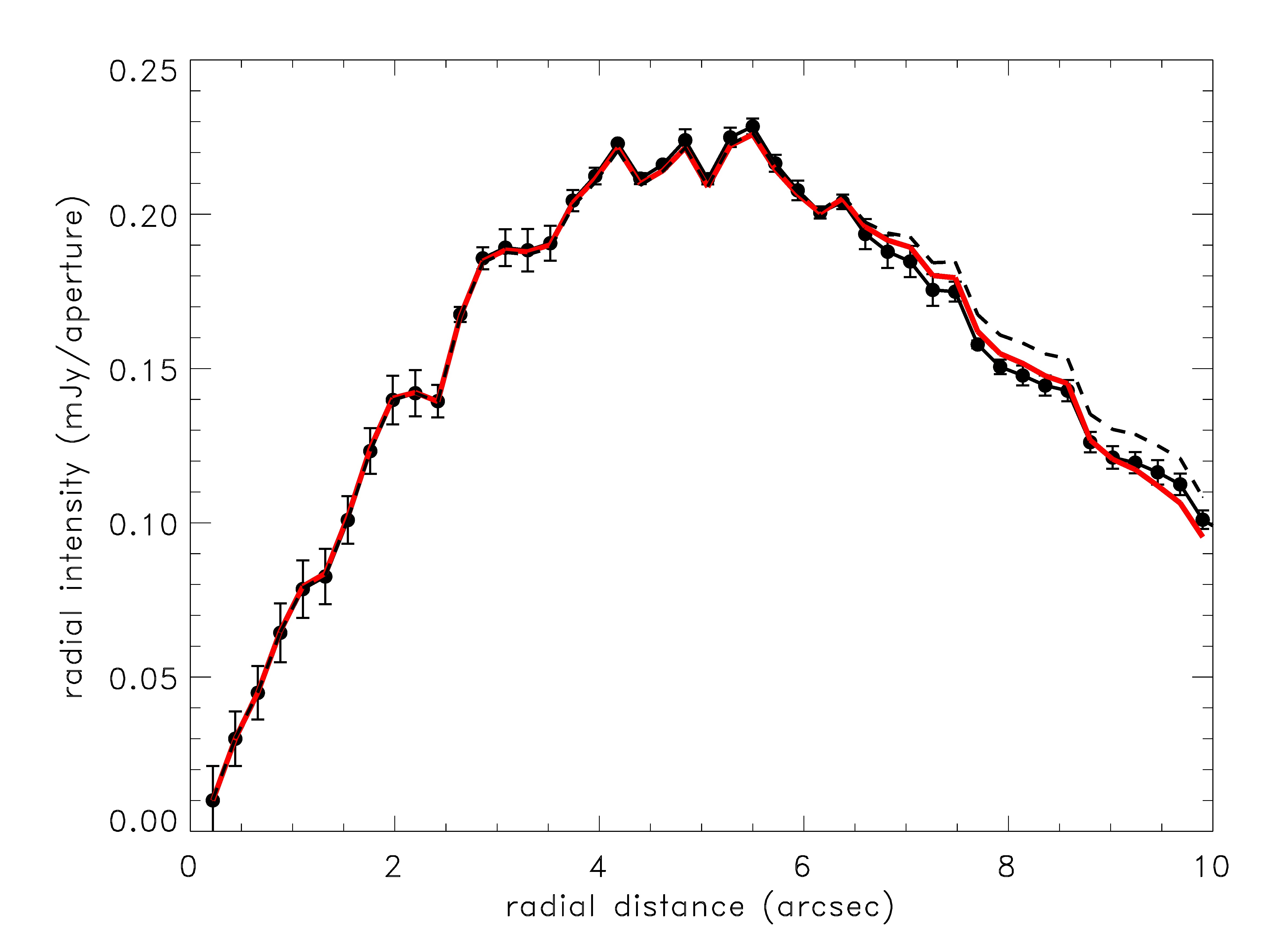}}
\caption{Radial intensity profiles in the three PACS band (70, 100 and
160\,$\mu$, from left to right). The observed profiles are presented by 
black dots, while colour curves represent the best fit model profiles 
according to Eq.~\ref{eq:coma} (see Table~\ref{table:knkc}). 
The corresponding radial intensity profiles of the respective PACS
PSFs are shown as dashed curves.}
\label{fig:intprof}
\end{figure*}


\section{Intensity profile \label{sec:photometric}}

The PSF width for PACS at 70, 100 and 160\,$\mu$m is 5.7, 7.5 and 11.7\arcsec, 
respectively. We detect a clear broadening of the comet PSF out to 10\arcsec{}
($\sim$50000\,km) at 70 and 100\,$\mu$m, allowing us to analyse the coma
structure at these wavelengths. 
At 160\,$\mu$m the coma structure is hardly resolvable due to the
wide PSF and the impact of the sky background (Fig~\ref{fig:intprof}). 
With the aim of trying to separate the nucleus and the coma, we first fitted the 
radially averaged intensity profiles in all PACS band with a two component model. 
One component corresponds to the unresolved nucleus, assuming a Dirac-delta at the 
intensity peak, while the another component describes the coma, with a radially 
decreasing surface brightness, characterized by the scale length $r_0$ and
the exponent $\gamma$:

\begin{equation}
F = k_{nucl}\delta(0) + {{k_{coma}} \bigg(1 + \bigg( {r\over{r_0}} \bigg)^{\gamma}} \bigg)^{-1}
\label{eq:coma}
\end{equation}
With these parameters we construct a 2D image that is convolved with the respective 
PSF of the fiducial calibration star $\gamma$\,Dra
to obtain a synthetic image and a corresponding radial intensity that is 
compared with the observed radial intensity profile. 
The best fit parameters are derived using a Levenberg-Marquardt fitter and
are presented in Table~\ref{table:knkc}. The best
fit coma profiles provides a nucleus contribution of 12.8, 8.2 and 7.5\,mJy at 
70, 100 and 160\,$\mu$m, respectively. In the red (160\,$\mu$m) band
this $k_n$\,=\,7.5\,mJy flux covers basically the total thermal emission
of the comet (see Table~\ref{table:knkc}), mainly due to the wide PSF. 
If these fluxes are considered as a flux originated from a 
solid nucleus, the size of the body can be estimated by thermal model 
calculations \citep{Mueller1998,Mueller2002} and results in a nucleus 
radius of r$_n$\,$\approx$\,11\,km, with a geometric albedo of p$_V$\,=\,5\%. 
Higher spatial resolution measurements 
obtained later, e.g. the Hubble Space Telescope measurements 
(Hubble Space Telescope, Li et al., 2014, in prep.\footnote{see the
Siding Spring Comet Workshop page at http://cometcampaign.org/workshop}) 
provided a nucleus size of r$_n$\,$<$1\,km indicating that the
thermal emission contribution we observe is probably not originated from a 
solid body, but from a spatially unresolved, compact dust coma, likely inside
the contact surface with the solar wind, as was observed in the case of  
29P/Schwassmann-–Wachmann with Herschel/PACS \citep{BM}. This is also supported
by the compact, unresolved coma at 160\,$\mu$m. 

We repeated the coma radial intensity profile fit without allowing any flux
contribution from the nucleus or compact dust coma($k_n$\,$\equiv$\,0). 
As the red (160\,$\mu$m) profile is very close to the red point source PSF
(see Fig.~\ref{fig:intprof}) we did not repeat this fit for this band. 
In the blue (70\,$\mu$m) and
green (100\,$\mu$m) bands the radial intensity profiles can be fitted almost equally 
well as in the $k_n$\,$\neq$\,0 case (see Table~\ref{table:knkc} 
and Fig.~\ref{fig:intprof}). The intensity decreases with the radial distance in 
both bands with $\gamma$\,$\approx$\,--2, slightly slower in the blue, 
indicating a relative excess of higher temperature / smaller particles at higher 
radial distances. The total fluxes derived from these fits are 
F$_{70}$\,=\,41$\pm$2\,mJy and F$_{100}$\,=\,26\,$\pm$2\,mJy 
in the blue and green bands, respectively. 
When we calculate the ratio of the blue and green intensity profiles 
(with 70$\mu$m intensity profiles convolved to the 100\,$\mu$m resolution)
in the inner parts of the coma { ($<$3\arcsec)} show a constant ratio that 
corresponds to the characteristic temperature of large grains at this
heliocentric distance, T\,$\approx$\,110\,K. At larger radial distances
{ ($>$5\arcsec{} of the observed radial intensity profile)}, however, 
the flux ratios and hence the characteristic temperatures increase, indicating
a relatively stronger presence of smaller grains in the outer regions. 
 
\begin{figure}[ht!]
\hskip -0.3cm
\hbox{\includegraphics[width=4.5cm]{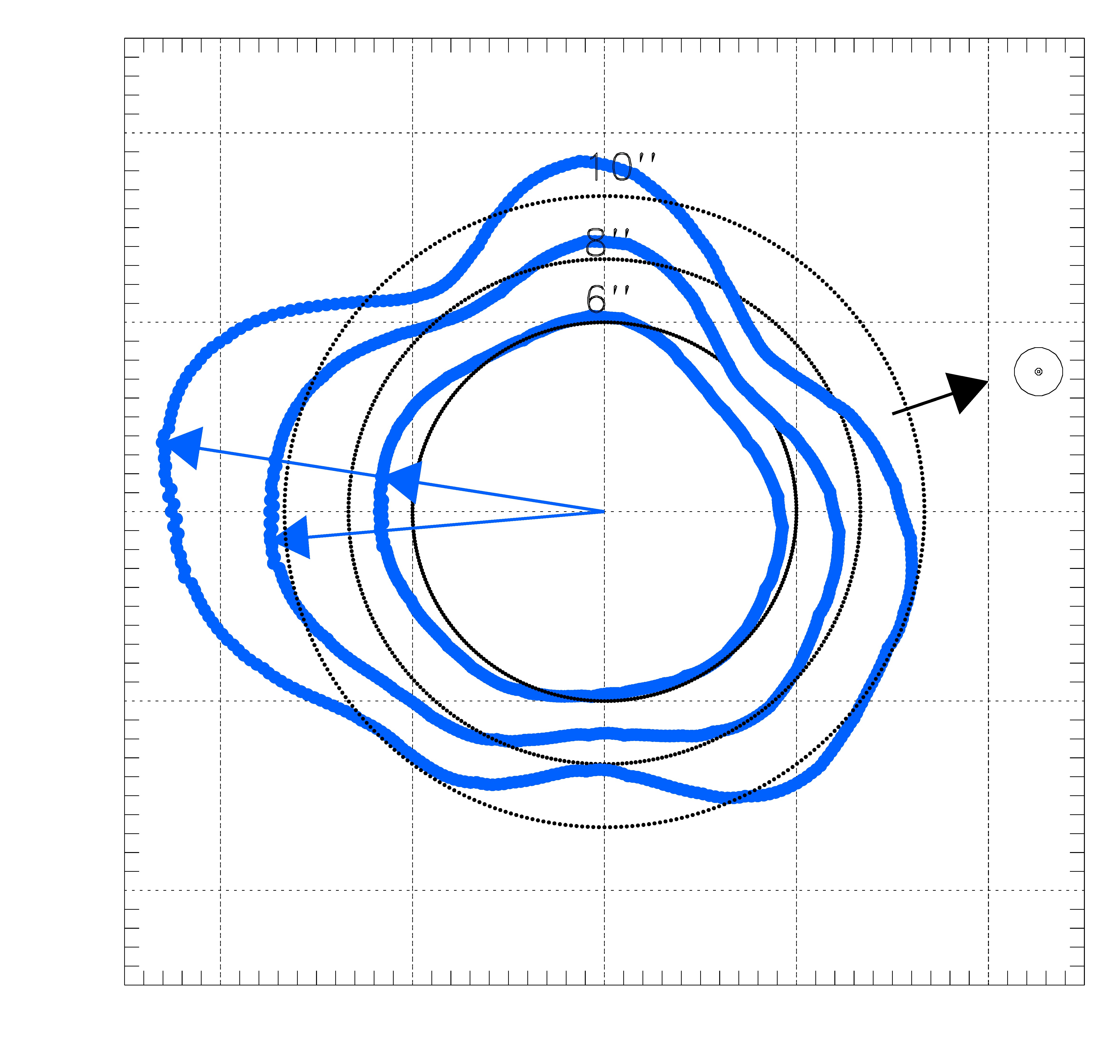}
\includegraphics[width=4.5cm]{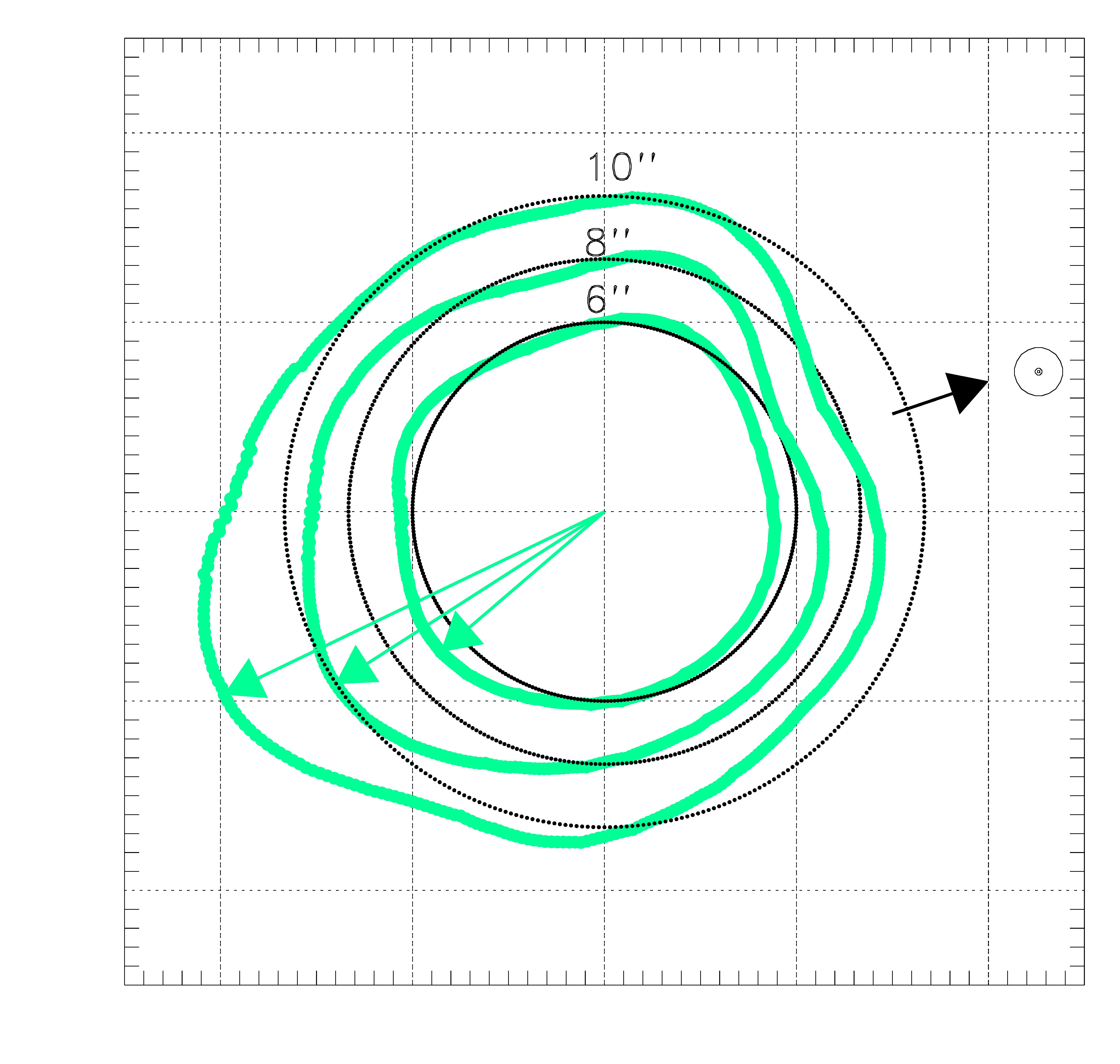}}
\caption{Dependence of integrated intensity on the position angle
within 6\arcsec, 8\arcsec and 10\arcsec from the intensity peak 
for the blue (left panel) and green (right panel) maps.
The contours show the intensity integrated in 30\degr~ 
segments w.r.t. the average value. The circles correspond to the mean
intensity at that specific radial distances, contours inside the circle
indicate intensity below the average while coutours outside the circle
indicate an excess. The coloured arrows show the the direction of the 
intensity maxima. The black arrow indicate the position angle of the Sun.}
\label{fig:elong}
\end{figure}

\begin{table}
\caption[]{Intensity profile fit results using Eq.~\ref{eq:coma}. 
The intensity
profiles were fitted both with and without including the contribution of a 
"nucleus"}. 
\begin{tabular}{c|rccc}
\hline
PACS & $k_n$ & $k_c$ & $\gamma$ & $r_0$  \\ 
band & (mJy) & (mJy/beam) &  & (arcsec)  \\ \hline
    & \multicolumn{4}{c}{ with nucleus}  \\
blue  & 12.8$\pm$0.7  & 0.13$\pm$0.02  &  1.89$\pm$0.04 &  2.51$\pm$0.14  \\
green &  8.2$\pm$0.8  & 0.12$\pm$0.03  &  2.15$\pm$0.05 &  2.54$\pm$0.16  \\
red   &  7.5$\pm$1.2  &   --   &  --   &  --     \\ \hline
      & \multicolumn{4}{c}{ without nucleus}  \\
blue  &  --   &  8.72$\pm$0.05   & 1.83$\pm$0.23 & 0.23$\pm$0.04    \\
green &  --   &  2.72$\pm$0.05   & 2.12$\pm$0.27 & 0.60$\pm$0.06    \\ \hline
\end{tabular}
\label{table:knkc}
\end{table}

On the 70 and 100\,$\mu$m images a slight elongation of the coma is
observed. On these images we derive the position angle of this feature using the intensity
integrated in a 30\degr{} cone out to a specific radial distance from the intensity
peak (see Fig.~\ref{fig:elong}). 
In the blue case the elongation contours are mostly affected 
by the strong tripod structure of the PACS 70\,$\mu$m PSF, overriding any 
other possible structure. In the green band, however, the tripod 
structure is less pronounced and in this band we obtained 
PA\,=\,$-139$\degr, $-147$\degr{} and $-154$\degr{} at the radial
distances of 6\arcsec, 8\arcsec and 10\arcsec. The position angle of the Sun 
was $\sim$19\degr{} at the time of the PACS observations.


\section{Simple dust production rate estimate}

From the observed thermal emission of the particles we can 
estimate the dust production rate in an $Af\rho$ manner 
\citep{AHearn_1984}. We calculate the $Af\rho$ parameter 
from the 70\,$\mu$m dust emission, as this emission is the least affected 
by the background. The thermal emission can be estimated as
\begin{equation}
F^{\nu}_{therm} = {{1-\overline{A}}\over{A(\alpha)}} \pi B_{\nu}(T) 
{{(Af\rho)}\over{\Delta^2}}\rho
\end{equation}
where { $\rho$ is the radial distance from the nucleus,}
$\overline{A}$ is the mean bolometric Bond albedo of the dust, 
A($\alpha$) is the phase angle dependent Bond albedo, 
$B_{\nu}(T)$ is the Planck function and we use the same 
parametrization as in \citet{Mommert}. The dust temperature at this 
heliocentric distance is estimated to be 108-116\,K for large 
($>$10\,$\mu$m) dust particles, with a slight dependence
on the particles type. { We adopt T\,=\,110\,K dust temperature and
$\rho$\,=\,50\,000\,km radial distance that corresponds to 10\arcsec{} apparent radial extension, 
resulting in a value of $Af\rho$\,=\,185$\pm$25\,cm. When this value is compared with
the $Af\rho$ values 
obtained from NEOWISE measurements at smaller 
heliocentric distances we found that the
increase of the activity was faster between 4 and 6.5\,au than would have been 
inferred from the \citep{Stevenson} data alone}. 
From this $Af\rho$ value we estimate the dust production rate: 
\begin{equation}
Q_{dust} = (Af\rho) {2\over{3}} {{{\rho_d}a{v_d}}\over{A_p}}
\end{equation}
where $\rho_d$ is the dust density, $a$ the dust grain radius, $v_d$ the escape 
velocity, and $A_p$ the geometric albedo of the dust particles, assuming a 
fixed grain size. 
We adopt a\,=\,15\,$\mu$m, $\rho_d$\,=\,1\,g\,cm$^{-3}$ and 
A$_p$\,=\,0.15 \citep{Kelley+Wooden}. For this particle size the dust velocity
is estimated as
\begin{equation} 
v_d = v_{ref} \Big( {{a}\over{1\,mm}} \Big)^{-0.5} 
\Big( {{r_h}\over{5\,au}} \Big)^{-1}
\label{eq:speed}
\end{equation}
Using v$_{ref}$\,=\,1.9\,m\,s$^{-1}$ \citep{Kelley} this results in
v$_d$\,=\,12\,m\,s$^{-1}$ for 15\,$\mu$m-sized grains
and Q$_{dust}$\,=\,1.5$\pm$0.5\,kg\,s$^{-1}$. We can compare these values
with the NEOWISE $Af\rho$-based estimates of Q$_{dust}$\,=\,11$\pm$4 and 
45$\pm$15\,kg\,s$^{-1}$ at 3.8 and 1.9\,au heliocentric distances, respectively 
\citep{Stevenson}.
{ These three Q$_{dust}$ estimates agree relatively well with a power-law 
distance scaling of ($r_h$/$r_0$)$^{q_Q}$, with an exponent of q$_Q$\,$\approx$\,--2.65, 
however, the activity started to fade right after the 1.9\,au measurement}.
Based on the assumptions above, we obtain a total coma dust mass of 3$\cdot$10$^8$\,kg.  

\section{Dust particle toy model}

The $Af\rho$ estimates presented above does not take into account 
that dust particles have a size and size-dependent velocity distribution.
To consider these effects, we constructed a more detailed emission model.
We assume that we have grains in the size range of 
a\,=\,$10^{-8}$--$10^{-2}$\,m and that the number of particles of a 
certain size scale as (a/a$_0$)$^q$, using
--1\,$\leq$\,q\,$\leq$\,--3. The equlibrium temperature of a grain of a 
certain size is calculated based on its optical properties (complex
refractory index) which are used to calculate the absorption efficiency 
Q$_{abs}$ at a specific wavelength and grain size, in the framework of the 
Mie-theory. Using these Q$_{abs}$ values we derive equlibirum temperatures for each
grain size. For details of the method see e.g. \citet{Jewitt}.
We used two types of grains: astrosilicates \citep{Draine} 
and glassy carbon particles \citep{Edoh}. 

In our model particles travel with a constant velocity -- the same as the
ejection speed -- that depends on their 
size and on the heliocentric distance at the time of their release
(see Eq.~\ref{eq:speed}). Using a similar scaling Farnham et al. 
(2014, in prep.$^1$) derived v$_{ref}$\,=\,0.42\,m\,s$^{-1}$ using HST observations. 
To allow a somewhat wider range, we let $v_{ref}$ 
vary in the range 0.25--1.0\,m\,s$^{-1}$ in our model. 
The comet was detected on PanSTARRS pre-discovery images in September 2012 
at $\sim$8\,au, but not on previous images from November-December 2011 at
$\sim$10.5\,au. The comet brightened more than 3$^{\rm m}$ between these two
epochs, while the increase for an inactive nucleus would only have been 0\fm6. 
This indicates that the activity started between these two dates
(Farnham et al., 2014, in prep.$^1$). 
{ To account for the different possible activity starting times we use 
a $t_0$ parameter in our model, the time spent from the 
activity switch on to the date of the Herschel/PACS observations.
$t_0$ has been chosen bewteen 1.6\,$\cdot$10$^7$\,s 
and 4.5$\cdot$10$^7$\,s, corresponding to the dates September 2012 and 
November 2011, respectively, to March 31, 2013. The time values (including $t_0$) 
in our model are converted to heliocentric distance using the orbit of the comet
obtained from the NASA/JPL Horizons\footnote{http://ssd.jpl.nasa.gov/?horizons}
database}. 
We also assume that the ejection
rate of the particles scales as $r_h^{-2}$, i.e. the activity increases
for smaller heliocentric distances.  

\begin{figure}
\includegraphics[width=9cm]{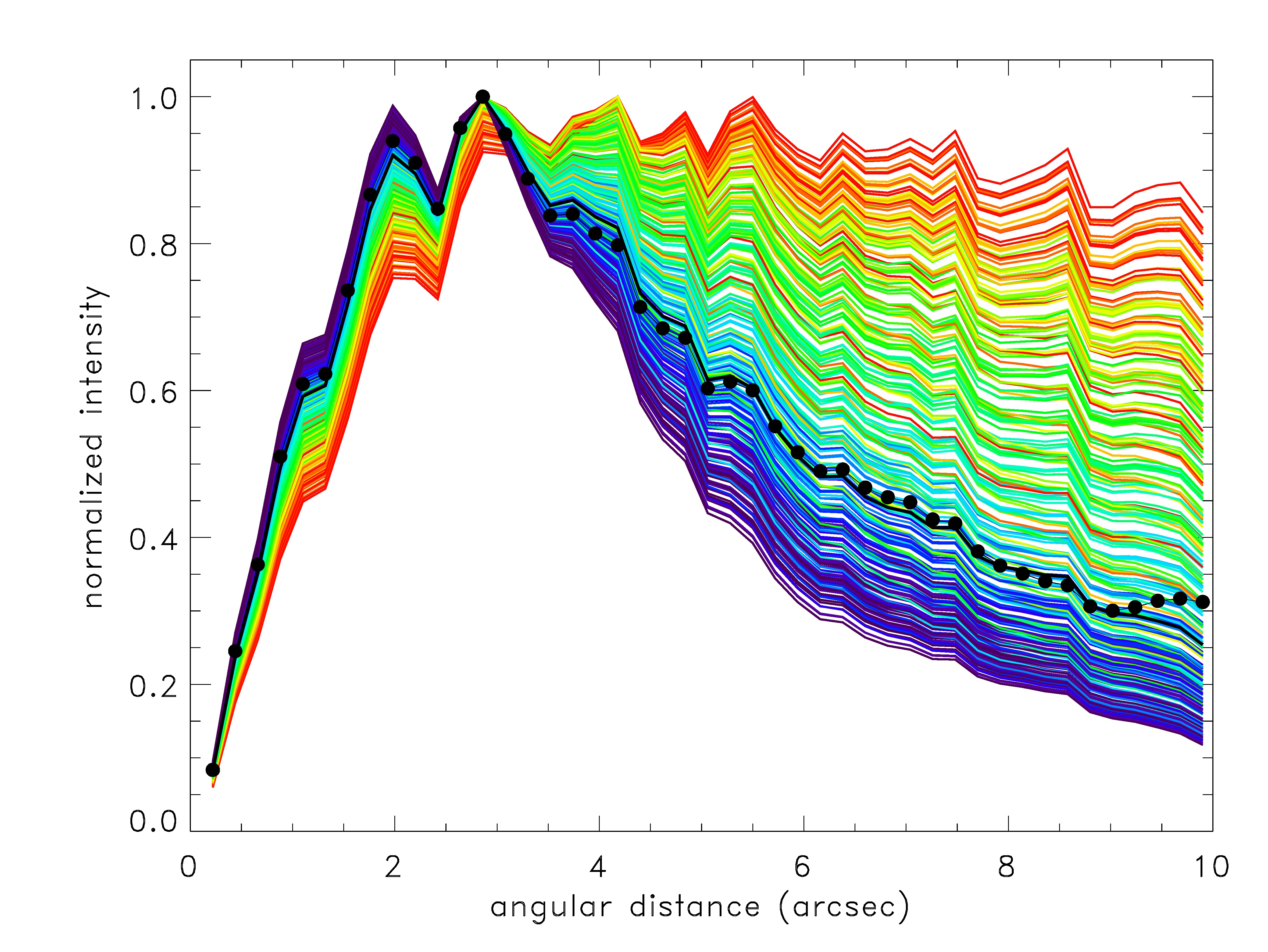}
\caption{Normalised intensity profiles of various model settings
(coloured curves). Curves with colours from blue to red correspond 
to size distribution parameters from q\,=\,--1.6 to --3.0.
The observed intensity profile is marked by black dots. 
The "best fit" (lowest $\chi^2$) curve (solid line) corresponds
to a model setting of t$_0$\,=\,1.6$\cdot$10$^7$\,s, v$_{ref}$\,=\,1.0\,m\,s$^{-1}$
and q\,=\,--2.0.}
\label{fig:toyintprof}
\end{figure}

Our model provides the radial distribution of particles of different sizes, 
and is used to calculate the 3D thermal emission and its 2D projection,
for a specific model setup. The 2D projection is then 
convolved with the respective Herschel/PACS beam to obtain a thermal 
emission distribution and radial intensity profile that is comparable
with the observed ones. We use the blue and green (70 and 100\,$\mu$m)
intensity profiles to determine which parameter set (t$_0$, v$_{ref}$ and q) 
provides the best fit to the observations. 
{ The goodness of fit is characterized by the $\chi^2$ values calculated from 
the observed and modelled intensity curves.}
An example is presented in Fig.~\ref{fig:toyintprof} where the
coloured curves correspond to different simulated intensity profiles at 70\,$\mu$m
using astrosilicates. 
{ The locations of differently coloured curves in the figure
suggests that models with steep size distributions (q\,$\approx$--3) 
provide a very extended radial intensity profile,
incompatible with the observed profile in the framework of our model. }
The best fit (lowest $\chi^2$) model solution corresponds 
to an activity onset time of 
{ t$_0$\,=\,1.6${_{-0.8}^{+1.6}}\cdot$10$^7$\,s (r$_h$\,=\,8\,au), 
a reference velocity of v$_{ref}$\,=\,1.0$\pm$0.3\,m\,s$^{-1}$ and 
q\,=\,--2.0$\pm$0.1. The errors of the best fit parameters are 
obtained requiring that the distribution of fit residuals of the two models are
not incompatible at the 2$\sigma$ confidence level.}
Our model profiles confirm the { relative} overabundance of large 
($\mu$m to mm sized) particles { compared with the generally assumed 
q\,$\approx$--3 size distribution. These particles are concentrated at smaller 
radial distances, as suggested by the grainsize-dependent ejection velocities.
Majority of the grains can be found at a distance of less than $\sim$1\arcsec, 
i.e. $\sim$5000\,km to the nucleus}. 
If carbon grains are used in the modelling instead of 
astrosilicates the model intensity profiles obtained are rather similar, 
with a maximum deviation of { $\sim$10\%} w.r.t. the astrosilicate intensity 
profile of the same parameter set { within the inner 10\arcsec{} of the coma. 
As presented in Fig.~\ref{fig:intprofdiff} the maximum deviation occurs for 
q\,$\approx$\,--3 models (red curves) at the outer parts of the region investigated
(5-10\arcsec). For the best fit q\,$\approx$\,--2 models (light blue curves) 
this deviation is much smaller, with a maximum deviation of $\sim$3\%, due to the higher
impact of larger, $\mu$m to mm sized grains, for which the optical properties
are less dependent on the grain type, i.e. either particle type
provides about the same best fit parameters in this case.}

\section{Summary}
\label{sec:summary}

Here we have reported on the thermal infrared observations of 
comet \css{} performed with the Herschel Space Observatory at a heliocentric 
distance of 6.48\,au. The comet showed an active
coma, detected in all PACS photometric bands (70, 100 and 160\,$\mu$m).
Using simple calculations based on the observed thermal emission we obtained 
a dust production rate of 1.5$\pm$0.5\,kg\,s$^{-1}$, and an $Af\rho$ value of
185$\pm$25\,cm, indicating a slow increase, unusual for an Oort cloud comet. 
The total dust mass of the coma is estimated
to be $\sim$3$\cdot$10$^8$\,kg. A more detailed dust grain model suggests that 
large grains are overabundant in the coma. 
Our model also indicates that the activity likely started close to
a heliocentric distance of 8\,au. 
  

\begin{acknowledgements}

Cs.~K. has been supported by the PECS grant \#\,4000109997/13/NL/KML of the 
HSO \& ESA, the K-104607 grant of the Hungarian Research Fund (OTKA) and the 
LP2012-31 "Lend\"ulet" grant of the Hungarian Academy of Sciences. We are indebted
to our referee for the useful comments. 

\end{acknowledgements}

{}

\appendix{\clearpage

\section{Appendix}

\begin{figure}[ht!]
\includegraphics[width=7.5cm]{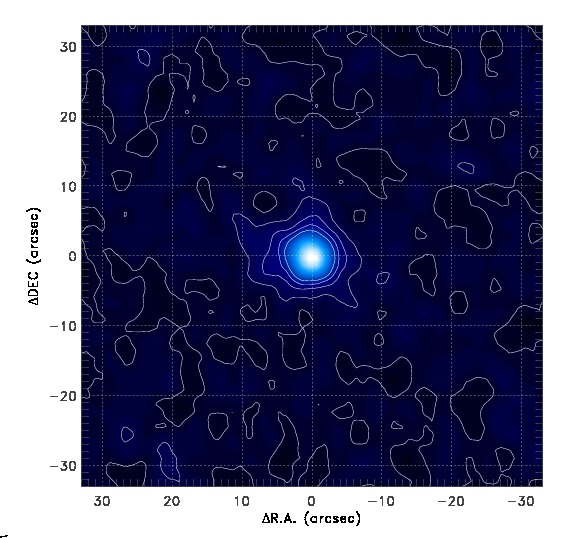}
\includegraphics[width=7.5cm]{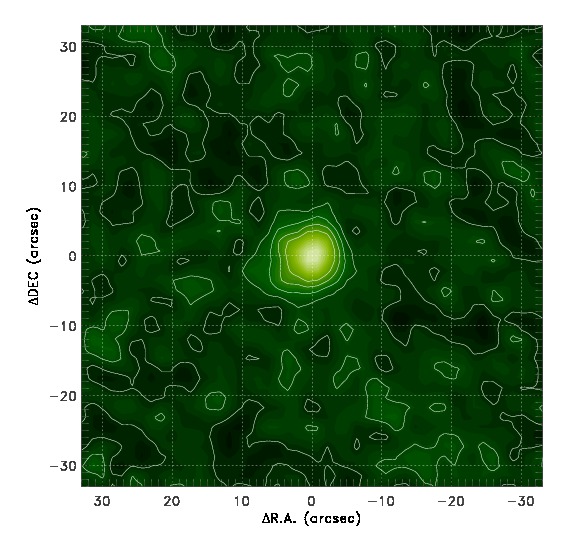}
\includegraphics[width=7.5cm]{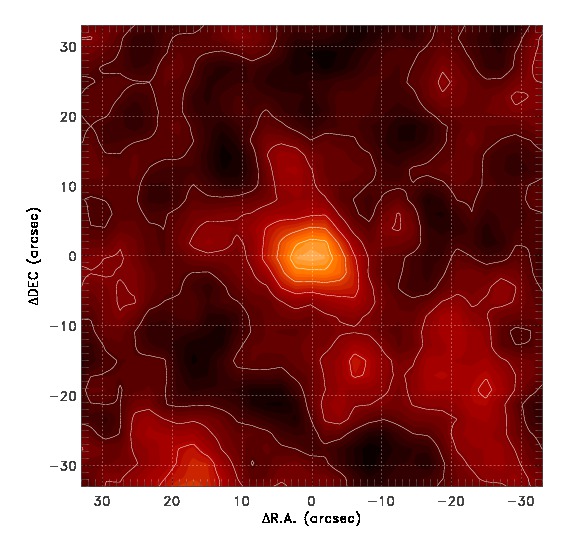}
\caption{Co-moving frame, co-added images of \css{} in the Herschel/PACS 
70 (top), 100 (middle) and 160\,$\mu$m (bottom) bands.  }
\label{fig:maps}
\end{figure}


\begin{figure}[ht!]
\includegraphics[width=8.5cm]{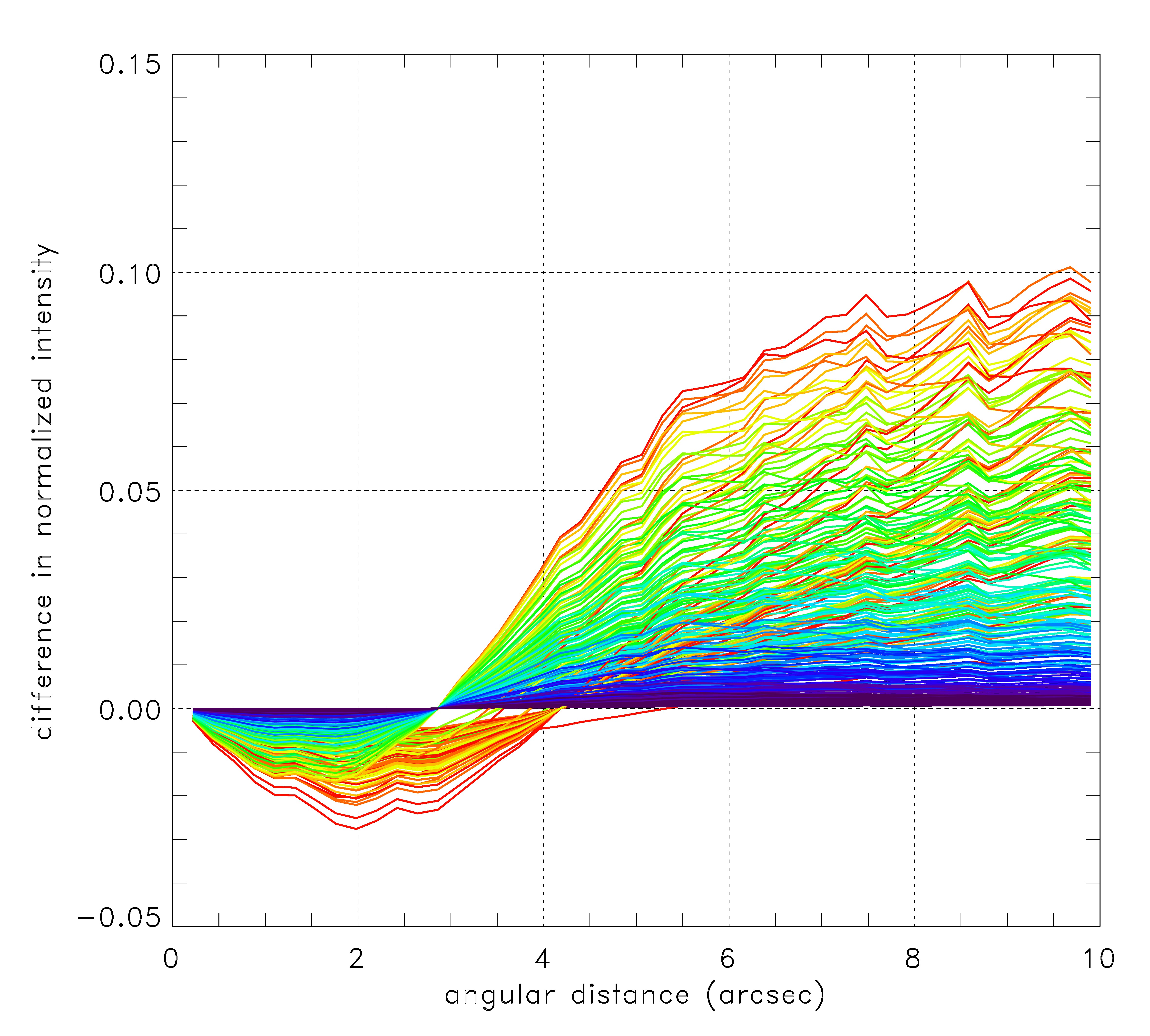}
\caption[]{Difference of the astrosilicate and carbon particle dust 
emission model intensity profile curves
using the same input parameters (t$_0$, v$_{ref}$ and $q$). 
Curves with colours from blue to red correspond 
to size distribution parameters from q\,=\,--1.6 to --3.0, as in 
Fig.~\ref{fig:toyintprof}}
\label{fig:intprofdiff}
\end{figure}


}

\end{document}